# Effect of 2 MeV $Fe^{3+}$ irradiation on Fe atom site population in a σ-phase Fe-Cr compound

*Stanisław M. Dubiel[1*] and Jan Żukrowski[1,2]*

[1]*AGH University of Science and Technology, Faculty of Physics and Applied Computer Science,* [2]*AGH University of Science and Technology, Academic Centre of Materials and Nanotechnology, al. A. Mickiewicza 30, PL-30-059 Kraków, Poland,*

A σ-$Fe_{54.5}Cr_{45.5}$ samples irradiated in vacuum with 2 MeV $Fe^{3+}$ ions at 300, 400, 475 and 700$^o$C to the maximum dose of 12.5 dpa were studied with the conversion electron Mössbauer spectroscopy (CEMS). The analysis of the room temperature CEMS spectra revealed an irradiation-induced redistribution of Fe atoms viz. their number on B and D sites decreased while on A, C and E sites increased. The degree of the redistribution was found to be proportional to the number of Fe atoms present on the lattice sites in the non-irradiated samples. The highest degree of the redistribution was revealed in the sample irradiated at 300$^o$C. No change in the site occupancy was found in the sample irradiated at 700$^o$C.

[*]Corresponding author: Stanislaw.Dubiel@fis.agh.edu.pl

Iron-chromium alloys have been of industrial importance because they constitute a major component of various brands of stainless steels. In particular, ferritic steels (FS), like ODS, and ferritic-martensitic (F-MS) ones, like EUROFER, are best examples of such materials. Their industrial and technological relevance follows from their very good swelling, high temperature corrosion and creep resistance properties [1,2]. Consequently, both FS as well as F-MS are regarded as appropriate construction materials for applications in new generations of nuclear power facilities such as generation IV fission reactors and fusion reactors as well as for other technologically important plants e.g. high power spallation targets [3–5]. In particular, such devices as fuel cladding, container of the spallation target or primary vessel are manufactured from these steels. These devices work at service not only at elevated temperatures, but also under long-term irradiation conditions. At elevated temperatures two phenomena may occur: (a) phase decomposition into Fe-rich ($\alpha$) and Cr-rich ($\alpha'$) phases, and (b) precipitation of a sigma ($\sigma$) phase. Both of them cause an enhancement of embrittlement, hence are highly undesired. The irradiation gives rise to an irradiation damage that can seriously deteriorate mechanical properties. On the lattice scale, the radiation causes lattice defects like vacancies, interstitials and dislocations. Consequently, a redistribution of Fe/Cr atoms occurs and can result in a short-range order (SRO), segregation or phase decomposition into $\alpha$ and $\alpha'$ phases. A better knowledge of the effects of irradiation on the useful properties of FS/F-MS and underlying mechanisms is an important issue as it may help to significantly improve properties of these materials, and to extend the operational life of devices constructed therefrom. The above-mentioned phenomena can be conveniently studied on Fe–Cr alloys that have been often used as model alloys for investigations of both physical and technological properties of stainless steels [6 and references therein]. In this Letter results concerning an effect of Fe-ions irradiation on a $\sigma$-Fe$_{54.5}$Cr$_{45.5}$ alloy are reported. The $\sigma$-phase has a tetragonal unit cell - space group $D^{14}_{4h}$ - P4$_2$/mnm - with 30 atoms distributed on five non-equivalent lattice sites – see Table 1.

Table 1

Characterization of the $\sigma$-phase in Fe-Cr. *WI* stands for the Wyckoff index, *CN* is a coordination number, *<d>$_{NN}$* is an average distance, in Å, to the nearest neighbors, *<IS>* is an average isomer shift in mm/s (relative to a Co(Rh) source). The values of *<d>$_{NN}$* and those of *<IS>* are taken from Ref. [7].

| Site | WI | CN | $<d>_{NN}$ | $<IS>$ |
|------|-----|----|-----------|--------|
| A | 2i | 2 | 2.506 | -0.38 |
| B | 4f | 4 | 2.702 | -0.03 |
| C | $8i_1$ | 8 | 2.655 | -0.16 |
| D | $8i_2$ | 8 | 2.572 | -0.35 |
| E | 8j | 8 | 2.640 | -0.27 |

Samples of σ-FeCr investigated in this study were prepared as described elsewhere [8]. For the irradiation, foils in form of ~20 mm rectangles and 0.2 mm thickness were used. They were irradiated at the JANNUS multi-ion beam irradiation platform at CEA, Saclay, France with 2.0 MeV $Fe^{3+}$-ions to the dose of $7.5 \cdot 10^{11}$ $Fe^{3+} \cdot cm^{-2}$, which is equivalent to the radiation damage of ~12.5 dpa in maximum, as calculated by the SRIM code. The full cascade method was used with the SRIM default values of threshold energies. No ion channeling is to be suspected due to polycrystalline structures of the samples. The irradiation area was circular and had a diameter of 20 mm. The irradiation was performed in vacuum at 300, 400, 475 and 700°C. Concentration and radiation damage (RD) profiles calculated with the SRIM code are shown in Fig.1. The samples were investigated by the conversion electron Mössbauer spectroscopy (CEMS), consequently a pre-surface zone of the samples with a thickness of ~0.3 μm – marked in Fig. 1 by a vertical stripe - was accessible to the measurements. It can be seen that the sample's volume measured with the applied technique was practically free of $Fe^{3+}$, hence the only effect of the irradiation can be of a ballistic origin. A level of the damage varies between ~3 and ~8.5 dpa, with an average of ~6 dpa. An example of CEMS spectra is presented in Fig. 2.

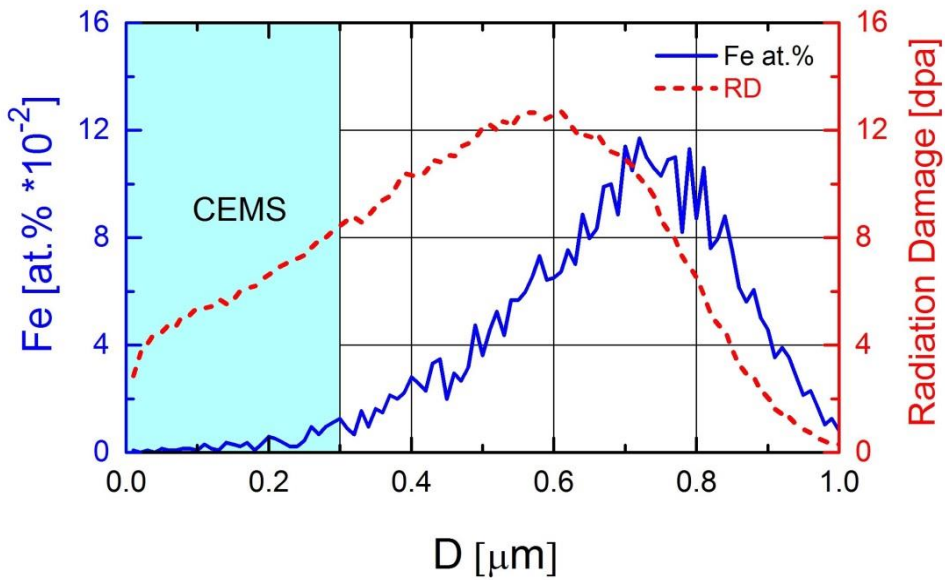

Fig. 1 $Fe^{3+}$ concentration and radiation damage profiles vs. depth, *D*, for the irradiated σ-FeCr sample.

A shape of the spectra recorded on the irradiated sides is, in general, similar to the one measured on the non-irradiated sides. This can be interpreted as a proof that the irradiation has not changed the crystallographic structure of the samples. Consequently, the spectra were analyzed in terms of five sub spectra (doublets) ascribed to each of the five lattice sites. Values of the isomer shift and of the quadrupole splitting of each doublet were fixed following the analysis described elsewhere [7], whereas linewidths and amplitudes of the lines were treated as free parameters. A relative spectral area of each sub spectrum can be regarded as a probability of finding Fe atoms on the lattice site represented by the sub spectrum. A comparison of such-calculated probabilities makes it possible to draw information on the effect of the irradiation on the population of Fe atoms on the five lattice sites. The upper panel of Fig. 3 illustrates the relative spectral area (RSA) as found from the analysis of the spectra measured on the non-irradiated sides.

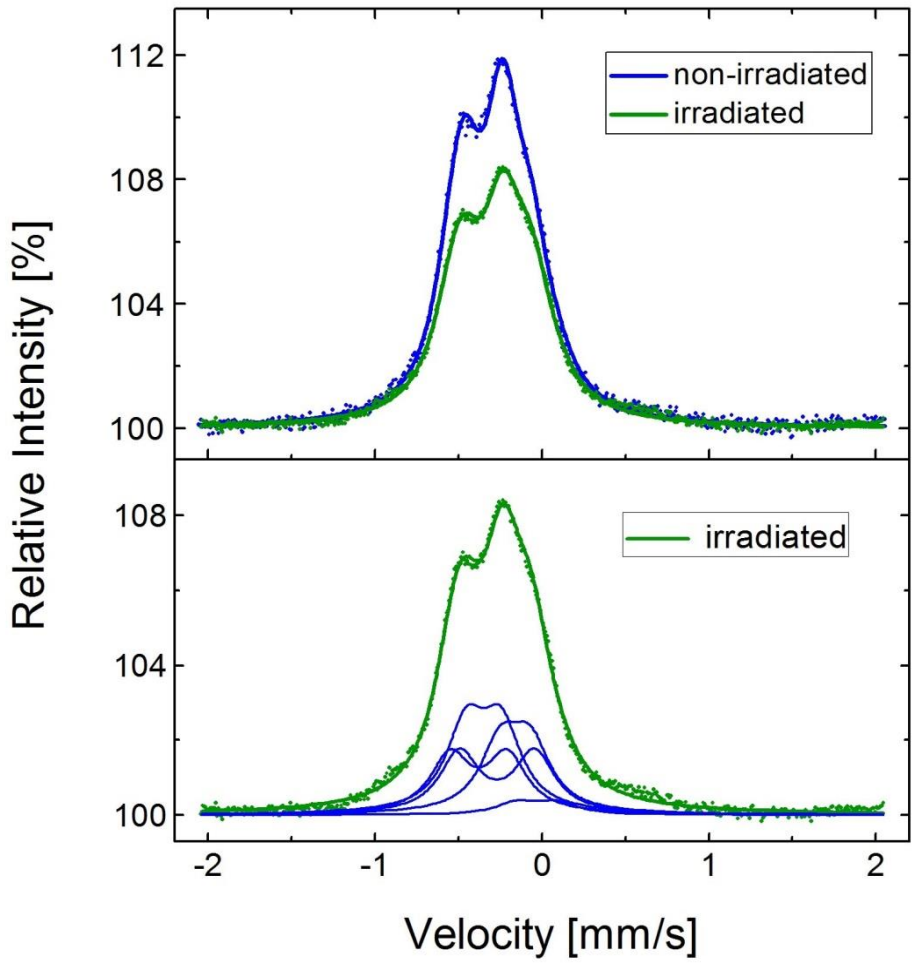

Fig. 2 CEMS spectra recorded at room temperature on non-irradiated and irradiated (300°C) sides of the σ-FeCr sample. Five sub spectra into which the measured spectrum was decomposed are indicated.

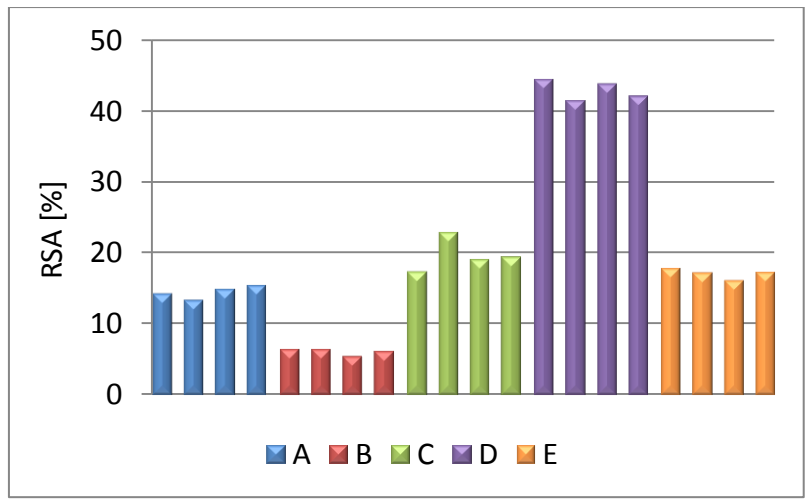

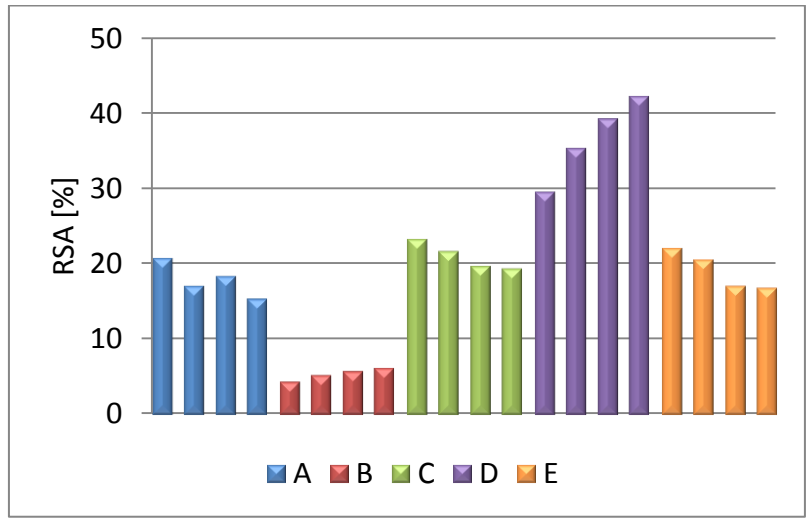

Fig. 3 Relative spectral area, RSA, as found from the spectra recorded on the: (upper panel) non-irradiated sides, and (lower panel) irradiated sides of the samples irradiated at 300, 400, 475 and 700°C (from left to right for each sub lattice).

It fallows that the RSA-values are characteristic of a given sub lattice, and they hardly depend on the temperature of irradiation.

Corresponding data determined for the irradiated sides are displayed in the upper panel of Fig. 3.

Here, the picture is different. Namely, one can distinguish two types of behavior viz. a decrease of RSA with temperature (sub lattices A, C, E) and its increase (B, D). To better visualize the behavior, a difference between the data shown in Fig. 3 was calculated and it can be seen in Fig. 4.

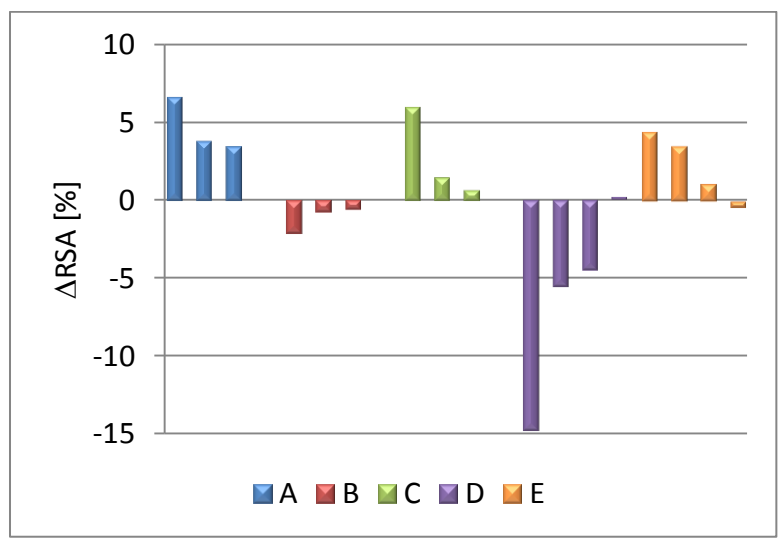

Fig. 4 Difference in the relative spectral area, ΔRSA, as found from the spectra recorded on the irradiated and non-irradiated sides of the samples irradiated at 300, 400, 475 and 700°C (from left to right for each sub lattice).

The data shown in Fig. 4 demonstrates how the population of Fe atoms on particular lattice sites changes upon the applied irradiation. It clearly depends both on the site and on the temperature. Namely, and an increase of the number of Fe atoms on the sites A, C, E is observed while their population on the sites B and D decreases. This means that the irradiation with $Fe^{3+}$ ions has caused an internal redistribution of Fe atoms. Interestingly, a degree of the redistribution strongly depends on the temperature of irradiation viz. it strongly decreases with the temperature: for 700°C no redistribution was found. Interestingly, this was the temperature at which a transformation of $\alpha$ into $\sigma$ in the studied samples was carried out [8]. The results clearly show that the five lattice sites are not equivalent as far as their stability against Fe-ion irradiation is concerned. In particular, a binding of Fe atoms occupying B and D sites is weaker than the one of Fe atoms residing on sites A, C, E. This issue concerns stability of $\sigma$ which is usually considered either in terms of electron concentration or atomic radius models [9]. Here, one is rather concerned with an internal stability of $\sigma$ i.e. changes in population at particular lattice sites. To tackle the issue one can consider, on one hand, a Fe atom charge-density (isomer shift) and, on the other hand, a radius of the nearest-neighbor (NN) shell. Average values of these parameters are displayed in Table 1. It is obvious that neither the charge-density nor the radius of the NN-shell can explain the observed redistribution of Fe atoms. Some light can be, however, shed on the issue when plotting the absolute value of the difference in the relative spectral area, |ΔRSA|, versus the number of Fe atoms on the lattice sites, N, see Fig. 5.

It is clear that the effect is proportional to the number of Fe atoms on a given site. This means that it has purely a statistical character. Yet, it does not explain why the Fe atoms are removed from the sites B and D.

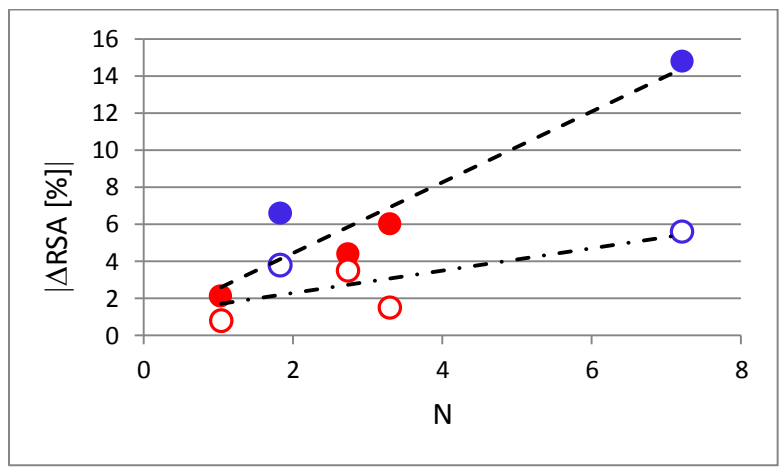

Fig. 5 The absolute value of the difference in the relative spectral area, |ΔRSA|, versus the number of Fe atoms on the lattice sites, N. Full symbols are for the irradiation at 300°C, and open ones for the one at 400°C. The lines are to guide the eye.

In summary, an effect of 2.0 MeV $Fe^{3+}$ irradiation at 300, 400, 475 and 700°C on a σ-$Fe_{53.8}Cr_{46.2}$ alloy was studied using CEMS Mössbauer spectroscopy. A clear cut evidence was found that upon the irradiation an internal redistribution of Fe atoms took place. Namely, a fraction of those originally occupying the sites B and D was replaced to A, C and E sites. The degree of the replacements was found to be proportional to the number of Fe atoms originally present on a given site. It also strongly depends on the temperature at which the irradiation was performed viz. the lower the temperature the higher the degree: at 700°C no change in the occupancy was revealed.

**Acknowledgements**

This work has been carried out within the framework of the EUROfusion Consortium and has received funding from the Euratom research and training programme 2014-2018 under grant agreement No 633053. The views and opinions expressed herein do not necessarily reflect those of the European Commission. It was also supported by The Ministry of Science and Higher Education, Warszawa, Poland and The AGH University of Science and Technology, Krakow, Poland. The irradiation was done at JANNUS-Saclay (Joint Accelerators for Nanoscience and NUclear Simulation), CEA, France and supported by the French Network EMIR. Mr. Yves Serruys is thanked for his expert assistance.